\documentclass[a4paper]{jpconf}
\usepackage{graphicx}
\begin{document}
\title{NNLO QCD corrections to event shape variables in electron positron
annihilation}

\author{A.\ Gehrmann-De Ridder$^1$, T.\ Gehrmann$^2$, E.W.N.\ Glover$^3$,\\
G.\ Heinrich$^4$}

\address{$^1$ Institute for Theoretical Physics, ETH, CH-8093 Z\"urich,
Switzerland}
\address{$^2$ Institut f\"ur Theoretische Physik,
Universit\"at Z\"urich, CH-8057 Z\"urich, Switzerland}
\address{$^3$ Institute of Particle Physics Phenomenology, 
Durham University, Durham, DH1 3LE, UK}
\address{$^4$ School of Physics, The University of Edinburgh, Edinburgh EH9 3JZ,
UK}

\vspace{-8cm}
{\noindent ZU-TH 24/07, IPPP/07/63, Edinburgh 2007/27}
\vspace{7.6cm}

\begin{abstract}
Precision studies of QCD at $e^+e^-$ colliders are based on measurements of 
event shapes and jet rates. To match the high experimental accuracy, 
theoretical predictions to next-to-next-to-leading order (NNLO) in QCD 
are needed for a reliable interpretation of the data. We report the
first calculation of NNLO corrections (${\cal O}(\alpha_s^3)$) to three-jet
production and related event shapes, and discuss their 
phenomenological impact. 
\end{abstract}

\section{Introduction}
Measurements at LEP and at earlier $e^+e^-$ colliders have 
helped to establish QCD as the theory of strong interactions by 
directly observing gluon radiation through three-jet production events. 
The LEP measurements of three-jet production and related event shape 
observables are of a very high
statistical precision. The extraction of $\alpha_s$ from these
data sets relies on a comparison of the data with theoretical predictions.
Comparing the different sources of error in this extraction,
one finds that the purely experimental error is negligible compared to
the theoretical uncertainty. There are two sources of theoretical
uncertainty: the theoretical description of the parton-to-hadron
transition (hadronisation uncertainty) and the uncertainty stemming from the 
truncation of the perturbative series at a certain order, as estimated by scale
variations (perturbative or scale uncertainty).  Although the precise
size of the hadronisation uncertainty is debatable and perhaps often
underestimated, it is certainly appropriate to consider the scale
uncertainty as the dominant source of theoretical error on the precise
determination of  $\alpha_s$ from three-jet observables. 

The three-jet rate and 
event shapes related to it can be expressed in 
perturbative QCD by dimensionless coefficients.
These coefficients depend either on the jet resolution parameter 
or on the event shape variable. Typically, one denotes these
coefficients by $A,B,C,\ldots$ at LO, NLO, NNLO, etc.

The perturbative expansion for the distribution of a generic observable
$O$ up to NNLO 
for renormalisation scale $\mu^2 = s$ and 
$\alpha_s\equiv \alpha_s(s)$  is given by
\begin{eqnarray}
\frac{1}{\sigma_0}\, \frac{d\sigma}{d O} &=& 
\left(\frac{\alpha_s}{2\pi}\right) \frac{d  A}{d O} +
\left(\frac{\alpha_s}{2\pi}\right)^2 \frac{d  B}{d O}
+ \left(\frac{\alpha_s}{2\pi}\right)^3 
\frac{d  C}{d O} \,.
\label{eq:NNLO}
\end{eqnarray}
normalised to the tree-level cross section for $e^+e^- \to q \bar q$.
It can easily be related to the physical observable 
$\frac{1}{\sigma_{\rm had}}\, \frac{d\sigma}{d O}$
using the known relation between $\sigma_{\rm{had}}$ and $\sigma_0$.

Ignoring the numerically negligible singlet contribution,
$A$, $B$ and $C$ depend only on the 
jet resolution parameter or the event shape variable 
under consideration, and are independent of electroweak 
couplings, centre-of-mass energy and renormalisation scale.

\begin{figure}[t]
\label{fig:delphi}
\begin{center}
\includegraphics[width=5cm]{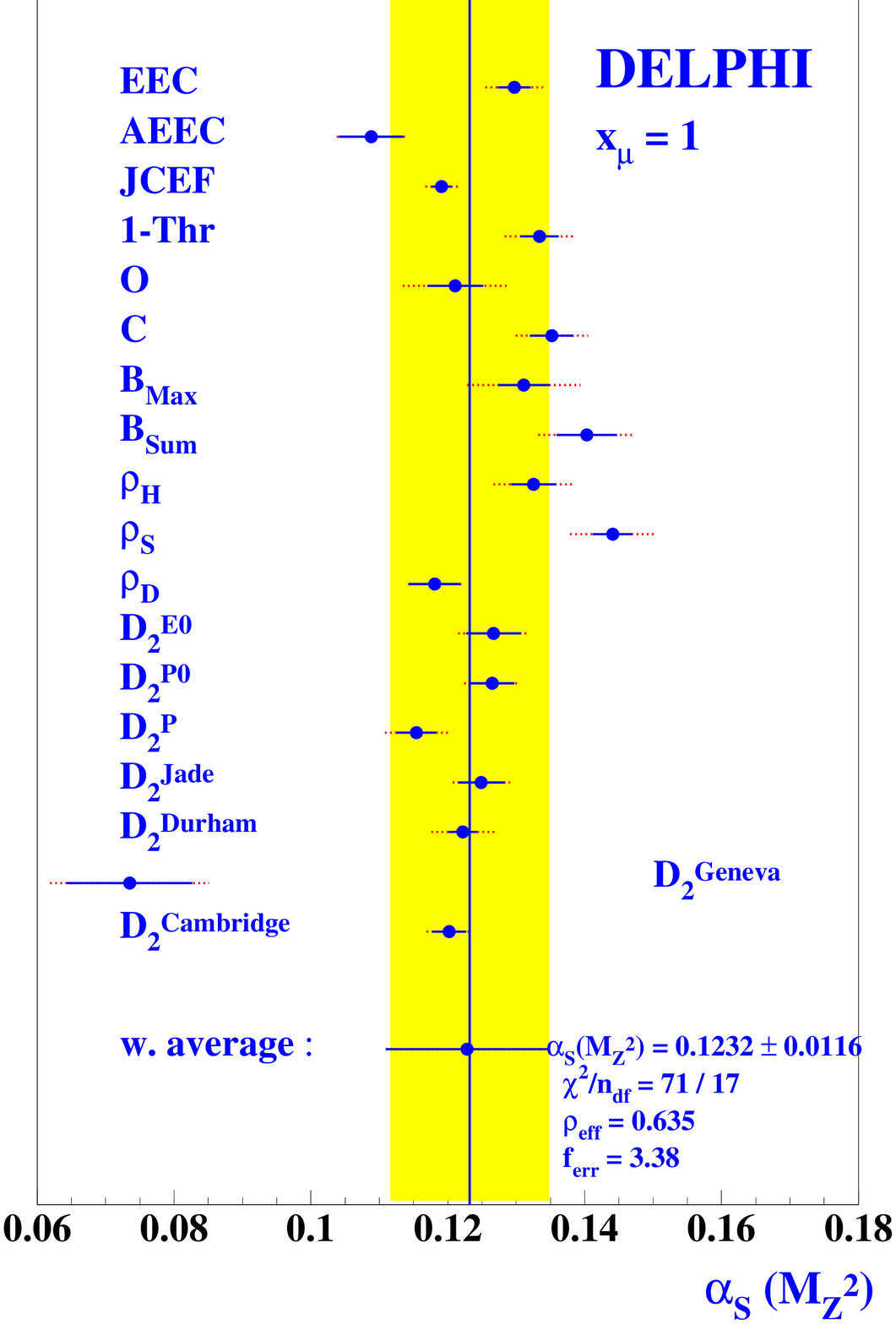}
\includegraphics[width=5cm]{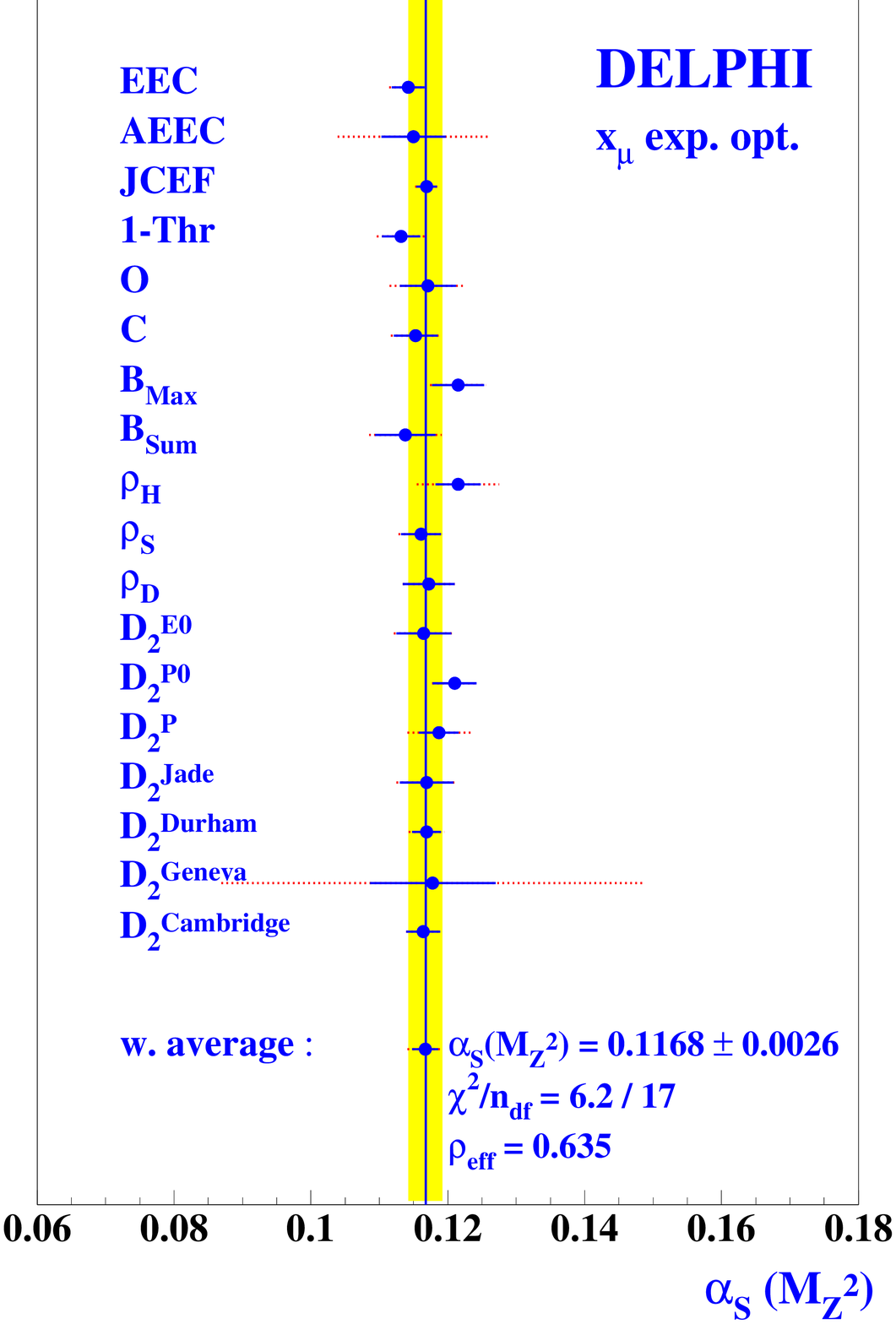}
\caption{Results of a NLO QCD fit to 18 hadronic observables
for (a) a fixed renormalisation scale $\mu = \sqrt s$ and (b) 
optimising the scale choice to fit the data.   The figures are
taken from Ref.~\cite{delphi}}
\end{center}
\end{figure}

QCD studies of event shape 
observables at LEP~\cite{expreview} based around the use of fixed-order
NLO parton-level event generator 
programs~\cite{event} have shown  
that the current error on $\alpha_s$ from these observables  
\cite{Bethke} is dominated by the theoretical uncertainty.
To illustrate this, the DELPHI collaboration made fits to 18 hadronic
observables (a) at fixed renormalisation scale, $\mu = \sqrt s$
and (b) by treating the renormalisation scale as a free parameter in the fit.
The results are displayed in Fig.~\ref{fig:delphi}, taken from Ref.~\cite{delphi}.
With a fixed scale (fig~\ref{fig:delphi}(a)), 
the value of the strong coupling extracted varies considerably amongst
leading to a value of $\alpha_s(M_Z) = 0.1232 \pm 0.0116$.   On the other hand,
fig~\ref{fig:delphi}(b) shows the result of 
letting the scale vary (and thereby estimating the uncalculated 
higher order corrections), and leads to a more consistent fit and a 
much smaller error  $\alpha_s(M_Z) = 0.1168 \pm 0.0026$.
Clearly, to improve the determination of $\alpha_{s}$, 
the calculation of the NNLO corrections to these
observables becomes mandatory.

\section{Thrust distribution}

\begin{figure}[t]
\includegraphics[width=4cm,angle=-90]{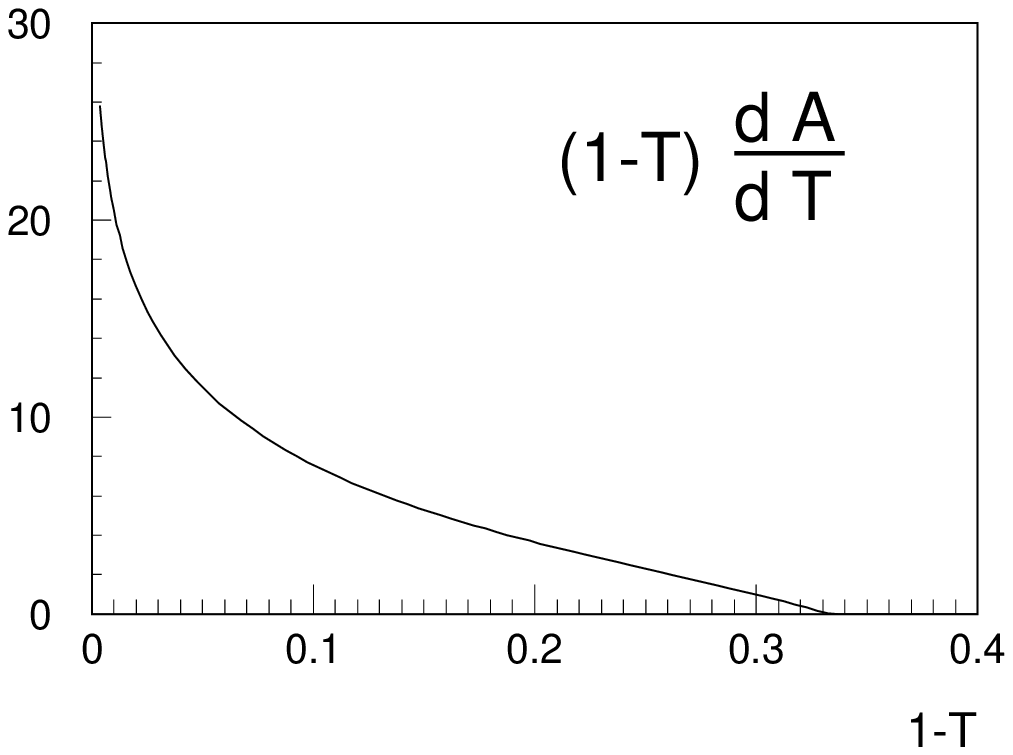}
\includegraphics[width=4cm,angle=-90]{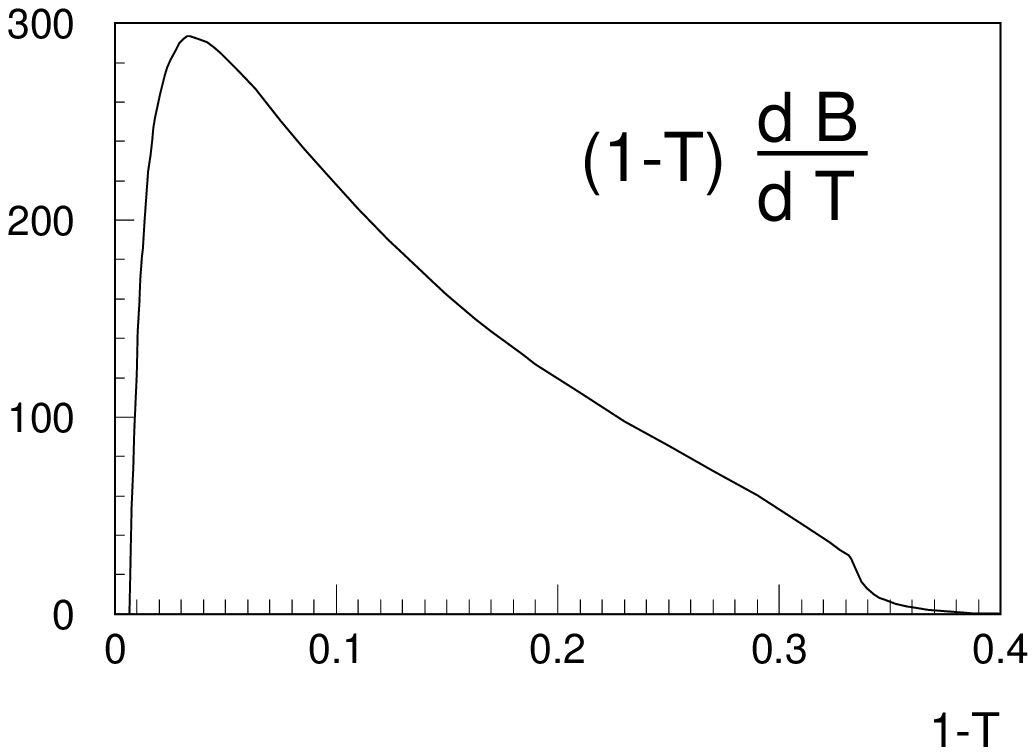}
\includegraphics[width=4cm,angle=-90]{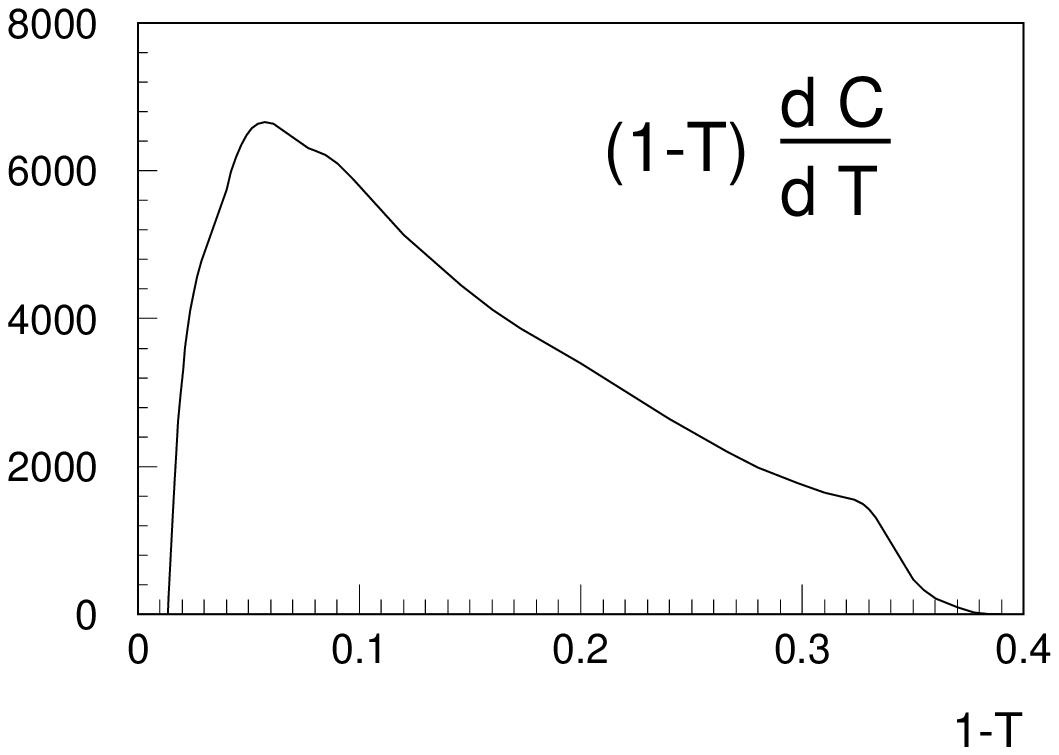}
\caption{Coefficients of the leading order, next-to-leading order and
next-to-next-to-leading order 
contributions to the thrust distribution.}
\label{fig:nlo}
\end{figure}

As an example, we focus on the thrust distribution, $O = T$.
The LO and NLO coefficients $A(T)$ and $B(T)$ are displayed for comparison  
in Figure~\ref{fig:nlo}. In the numerical evaluation, we use $M_Z= 91.1876$~GeV and $\alpha_s(M_Z)=
0.1189$~\cite{Bethke}.

The NNLO coefficient $C(T)$ has been recently
computed~\cite{ourT}.  It is based on two-loop $\gamma^* \to q\bar q g$ 
matrix elements~\cite{3jme,muw2}, 
one-loop four-parton matrix elements~\cite{V4p} and tree-level
five-parton matrix elements~\cite{tree5p}.

The two-loop $\gamma^* \to q\bar q g$ matrix elements were derived 
in~\cite{3jme} by reducing all relevant Feynman integrals to a small 
set of master integrals using integration-by-parts~\cite{ibp} and 
Lorentz invariance~\cite{gr} identities. The master integrals~\cite{3jmaster} were 
computed from their differential equations~\cite{gr} and expressed 
analytically
in terms of one- and two-dimensional harmonic polylogarithms~\cite{hpl}. 

The one-loop four-parton matrix elements relevant here~\cite{V4p} were 
originally derived in the context of NLO corrections to four-jet 
production and related event shapes~\cite{fourjetprog,cullen}. 

The four-parton and five-parton contributions to three-jet-like final 
states at NNLO contain infrared real radiation singularities, which 
have to be extracted and combined with the 
infrared singularities 
present in the virtual three-parton and four-parton contributions to 
yield a finite result. In our case, this is accomplished by 
introducing antenna subtraction functions~\cite{ourant,our2j}, 
which encapsulate 
all singular limits due to the 
 emission of one or two unresolved partons between two colour-connected hard
partons, and are sufficiently simple to 
be integrated analytically~\cite{ggh}. 

The resulting numerical programme, {\tt EERAD3}, yields the full kinematical 
information on a given multi-parton final state. It 
can thus be used to compute any 
infrared-safe observable related to three-particle final states at 
${\cal O}(\alpha_s^3)$. The NNLO coefficient $C(T)$ for the 
thrust distribution obtained with {\tt EERAD3} is shown in
Fig.~\ref{fig:nlo}~\cite{ourT}.

\begin{figure}[t]
\begin{center}
\includegraphics[width=5cm,angle=-90]{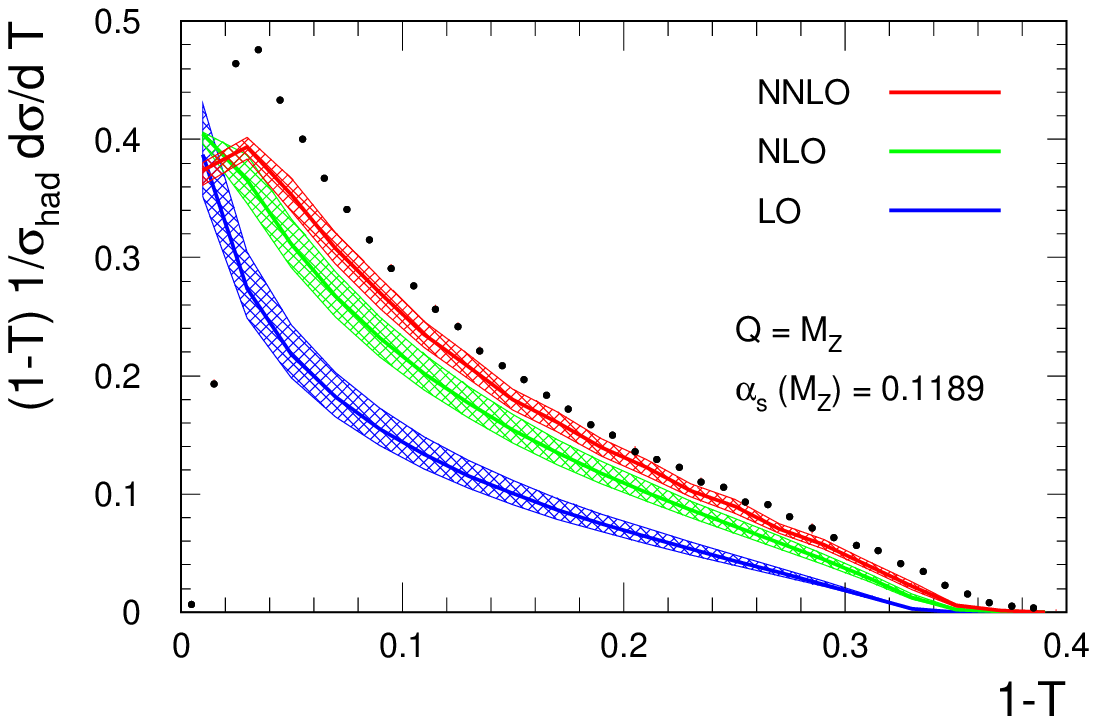}
\includegraphics[width=5cm,angle=-90]{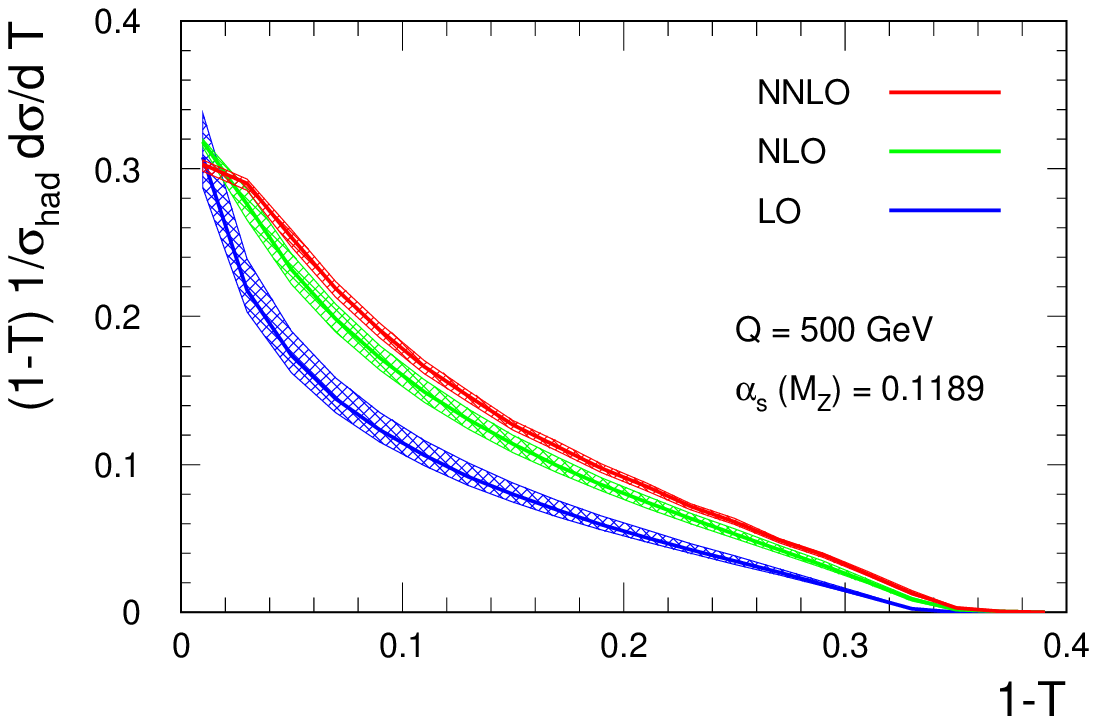}
\caption{Thrust distribution at LEP and at
the ILC with $Q= 500$~GeV.\label{fig:thrust}}
\end{center}
\end{figure}

Figure~\ref{fig:thrust} displays the perturbative 
expression for the thrust distribution at LO, NLO and NNLO, evaluated 
for LEP and ILC energies. The error band indicates the variation of the 
prediction under shifts of the renormalisation scale 
in the range $\mu \in [Q/2;2\,Q]$ around the $e^+e^-$ centre-of-mass 
energy $Q$.

It can be seen that even at linear collider energies,
inclusion of the NNLO corrections 
enhances the thrust distribution by around 10\% over the 
 range $0.03 < (1-T) < 0.33$, where relative scale uncertainty is
 reduced by about 30\% between NLO and NNLO. 
Outside this range, one does not expect the 
perturbative fixed-order prediction to yield reliable results. 
For $(1-T)\to 0$, 
the convergence of the perturbative series 
is spoilt by powers of
logarithms $\ln(1-T)$ appearing in higher perturbative orders, 
thus necessitating an all-order resummation of these logarithmic 
terms~\cite{ctwt,ctw}, and a matching of fixed-order and resummed 
predictions~\cite{hasko}.

The perturbative parton-level
prediction is compared with the hadron-level data from
the ALEPH collaboration~\cite{aleph}
 in Figure~\ref{fig:thrust}. 
 The shape and
normalisation of the parton level NNLO prediction agrees better with the
data than at NLO.
We also see that the NNLO corrections account
for approximately half of the difference between the parton level NLO
prediction and the data.  

\section{Conclusions}
We developed 
a numerical programme which can compute any infrared-safe observable
through to ${\cal O}(\alpha_s^3)$, which we applied here to determine the 
NNLO corrections to the thrust distribution. These corrections are moderate,
indicating the convergence of the perturbative expansion. Their inclusion 
results in a considerable reduction of the theoretical error on the 
thrust distribution and will allow a 
significantly  improved determination of 
the strong coupling constant from jet observables from existing LEP data.

\section*{Acknowledgements}
This research was supported in part by the Swiss National Science Foundation
(SNF) under contract 200020-109162, 
 by the UK Science and Technology Facilities Council and by the European 
Commission under contract MRTN-2006-035505 (Heptools).




%



\end{document}